\documentclass[preprint,12pt, a4paper]{elsarticle}

\usepackage{amssymb}
\usepackage{url}

\usepackage{lineno}

\journal{SoftwareX}

\begin{document}

\begin{frontmatter}



\title{BinarySDG: binary sensor data generation with R}

\author[label1,label2]{M. Piangerelli}
\author[label1]{G. Rocchetti}
\author[label1]{A. Liscio}
\author[label2]{R. De Leone}
\address[label1]{School of Science and Technologies, Computer Science Division, University of Camerino}
\address[label2]{School of Science and Technologies, Math Division, University of Camerino}
\begin{abstract}
The scarcity of Smart Home data is still a pretty big problem, and in a world where the size of a dataset can often make the difference between a poor performance and a good performance for problems related to machine learning projects, this needs to be resolved. But whereas the problem of retrieving real data can’t really be resolved, as most of the time the process of installing sensors and retrieving data can be found to be really expensive and time-consuming, we need to find a faster and easier solution, which is where synthetic data comes in. Here we propose BinarySDG (Binary Synthetic Data Generator) as a flexible and easy way to generate synthetic data for binary sensors.

\end{abstract}


\end{frontmatter}



\section{Motivation}
\label{}

Our need for synthetic data derived from the necessity of having data in a period of time shorter than the one needed by the process of bureaucracy, sensors retrieval,  sensors installing and data retrieval itself, which can very often be found to be difficult, expensive and time-consuming, and that’s where synthetic data comes in.

When it comes to synthetic data generation, it comes from itself that one already has in mind the data he’s trying to generate, or already has a small dataset to use as a sample.
With this being said, in general, it would be pointless to use synthetic data for tasks as features extraction or data analysis, as the generated data would only be just a derivation of an already existing structure.

What we needed was a dataset that could represent the behavior of a person living in an apartment, so that we could get to recognize a pattern of his daily routine, and from here trying to detect any kind of anomaly in the behavior. This automatically split into two different needs: the first was creating a dataset with sensors similar to the ones that would’ve been installed in the real problem application; the second was the necessity to have different behaviors of different subjects so that we could work on a wider spectrum of observations.

The internet can most of the times be the solution to problems related to machine learning projects but this one wasn’t the case; even though we found a few solutions regarding the generation of synthetic data for sensors, none of them could really fit in our problem domain: most of these solutions rely on activity recognition for every sequence of sensors activations, which is a feature that we couldn’t provide and wouldn’t represent our real project application; but the main problem was that none of these solutions could provide us a way to manage different behaviors and anomalous patterns generation.

Some bibliography that the web could provide us was the following:
some of the earliest efforts to simulate human behaviour through sensors data used mathematical models such as Markov chains and Petri networks to recreate a psychological
and physiological pattern \cite{physiological}, some later work by used again Markov chains but combining them with a Poisson spike generator to generate timestamps for the sensors \cite{pervasivespaces1} \cite{pervasivespaces2}.
Some more recent work used generative adversarial networks (GANs) to generate data and employing an adversary model to determine whether the generated samples could be mistaken with the real data \cite{gans}.

We also found two open source softwares that we managed to retrieve and analyze from the web: SynSys \cite{synsys} and SHGen \cite{shgen}, but none of them could provide us the solution we were looking for.

With this being said, even though the web could offer a few different possible solutions to the synthetic data generation problem, the most of them would either need a pretty good knowledge of the data analysis field or particular characteristics or size of the datasets from where to start generating data, but we wanted to create a more flexible yet easy way of generating data.

What we offer is an innovative and hybrid solution that can satisfy both a user with no ready dataset but a general behavior and structure in mind, and one that simply wants to multiply the dataset he already has by generating similar data.
On top of this, we offer an efficient anomaly generator, where the user can decide the number, dates, and duration of the anomalies he wants to generate in the dataset.

\section{Software description}
\label{}

BinarySDG permits the creation of binary sensor data through the use of the Markov Chain, a stochastic process which moves among the elements of a set X in the following manner: when at $x \in X$, the next position is chosen according to a fixed probability distribution P(x,·) depending only on x. This behavior is called Markov Property: the conditional probability of proceeding from state x to state y is the same, no matter what sequence $x_0,x_1,...,x_{t-1}$ of states precedes the current state x \cite{markovchains}. We can then imagine this process as a state machine where each element of a set corresponds to a state and each edge that links a state to another one has an assigned probability which represents the probability of moving from the origin state to the other. Transition rules are given by the Transition Matrix: given the transition matrix M, each element Mij represents the probability of moving from state x to state y.

\begin{figure}
  \includegraphics[width=\linewidth]{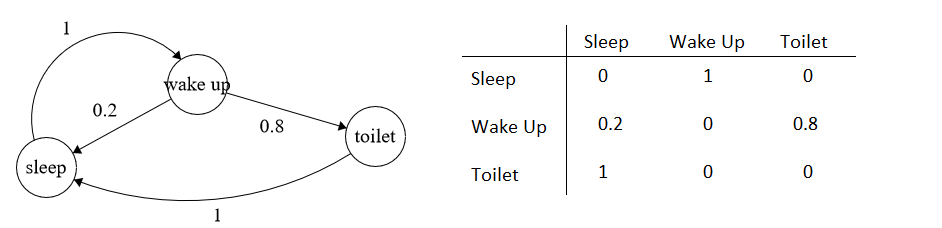}
  \caption{An example of Markov Chain with relative Transition Matrix.}
  \label{fig:mc}
\end{figure}

BinarySDG generates data using multiple Markov Chains based on the hour of the day, so it is possible to define different behaviors during the day.

\subsection{Software Architecture}
\label{}

BinarySDG uses the R programming language, it depends on the functionality of markovchain \cite{mc_package} and lubridate \cite{lubridate} libraries and it follows a procedural programming paradigm with the definition of S3 objects.

\subsection{Software Functionalities}
\label{}

BinarySDG provides the following methods for generating binary sensor data:

\begin{table}[!h]
\begin{tabular}{|p{6.5cm}|p{6.5cm}|}
\hline
\textbf{Method} & \textbf{Description} \\
\hline
generator\_from\_matrix() & Creates an S3 object of class generator\_from\_matrix \\
\hline
generator\_from\_dataset() & Creates an S3 object of class generator\_from\_dataset \\
\hline
start() & Starts generating data from a generator object \\
\hline
\end{tabular}
\caption{BinarySDG methods.}
\label{} 
\end{table}

\subsubsection{Generator From Matrix}
\label{}

The generator\_from\_matrix class is used to generate sensor data given in input the transition matrices needed to define the subject’s behavior and each state defines an activity.
A generator\_from\_matrix object takes five CSV files as input: four are transition matrices for building Markov Chains (morning, afternoon, evening, night) and the last one is a file containing information for each activity (type of activity, corresponding sensor, mean and standard deviation of the activity’s duration).

\subsubsection{Generator From Dataset}
\label{}

The generator\_from\_dataset class is used to generate sensor data given in input a sample dataset, from where the generator will abstract the main characteristics that will be used to define a pattern that can simulate the dataset’s core structure, such as mean and standard deviation of the duration of the sensors of which the input dataset is composed. A generator\_from\_dataset object takes a CSV file containing the dataset and an integer corresponding to the time interval (in hours) that divides the day (for example, with interval = 3, then we will have 24/3 = 8 Markov Chains).

\subsubsection{Start}
\label{}

The start method requires two mandatory parameters: a generator (created as described upon) and an integer representing the number of days to generate.
The method first takes all Markov Chains and duration information from the generator object, then the main flow of the method starts. It selects the Markov Chain corresponding to the current time and computes the next activity/sensor (based on the type of generator) to insert into the sequence. Then it checks if the selected item is valid in that point of the sequence (for example, if television is already switched on, it can not insert a tv\_ON event): if it is a valid event, the method computes the duration from a normal distribution based on the event’s mean and standard deviation taken from the generator. The event is appended to the result, the current time is increased by the event’s duration and the method repeats the process until the current time reaches the user’s desired number of days.
\begin{itemize}
  \item unusual activity anomaly: the method selects a Markov Chain corresponding to another random period of the day (except the true one);
  \item unusual duration anomaly (long or short): the mean of the event’s duration is changed into the boundary of outlier values (3 times the interquartile range of the distribution);
  \item both.
\end{itemize}

\section{Illustrative Examples}
\label{}

As a demonstration of the functionality of BinarySDG here we provide two different examples: the first example shows the procedure and the result of generating one week of synthetic data using the generator\_from\_matrix object (i.e. giving in input transition matrices), while the second one uses the generator\_from\_dataset object.
In both the examples the generated dataset goes from 1970-01-01 to 1970-01-07, with a constant anomaly generation from 1970-01-04 00:00:00 to 1970-01-06 23:59:59.

In line 2 we create the generator\_from\_matrix object, giving in input the four transition matrices files (one for each interval of the day) and the activities information file, where the user can define the name, type, relative sensor, mean durations and standard deviations of the durations of the activities.

\begin{figure}[!ht]
  \includegraphics[width=\linewidth]{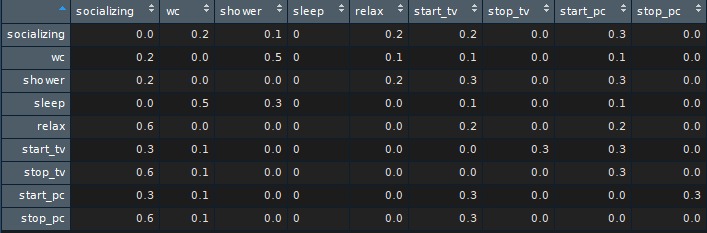}
  \caption{An example of morning transition matrix.}
  \label{fig:transitionmatrix}
\end{figure}

\begin{figure}[!ht]
  \includegraphics[width=\linewidth]{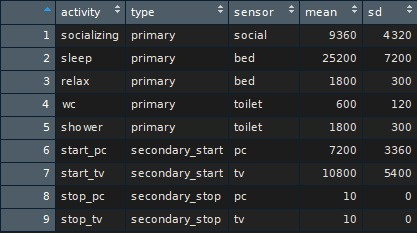}
  \caption{An example of activities information.}
  \label{fig:activities}
\end{figure}

Even though the thought of creating these input files might seem pretty scary at first, we can ensure that, with the software being pretty quick in completing the task, it really shouldn’t be.
We witnessed that firstly setting up some raw matrices and then running the package several times and tweaking up the parameters each time ended up to be much easier than we thought.

\begin{figure}[!ht]
  \includegraphics[width=\linewidth]{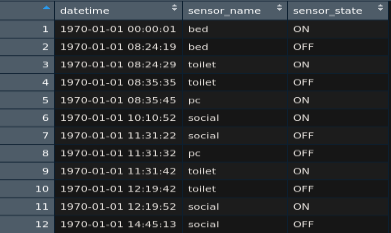}
  \caption{An example of input dataset.}
  \label{fig:dataset}
\end{figure}

In line 7 we create the generator\_from\_dataset object, giving in input the sample dataset file and setting a value of 3 as interval parameter, meaning that the input dataset will be analyzed splitting it in intervals of 3 hours each, for a total of 8 intervals a day, and the output dataset will be generated in the same manner.
Lines 4 and 8 show the call of the function start, used to generate the data.

\begin{figure}[!ht]
  \includegraphics[width=\linewidth]{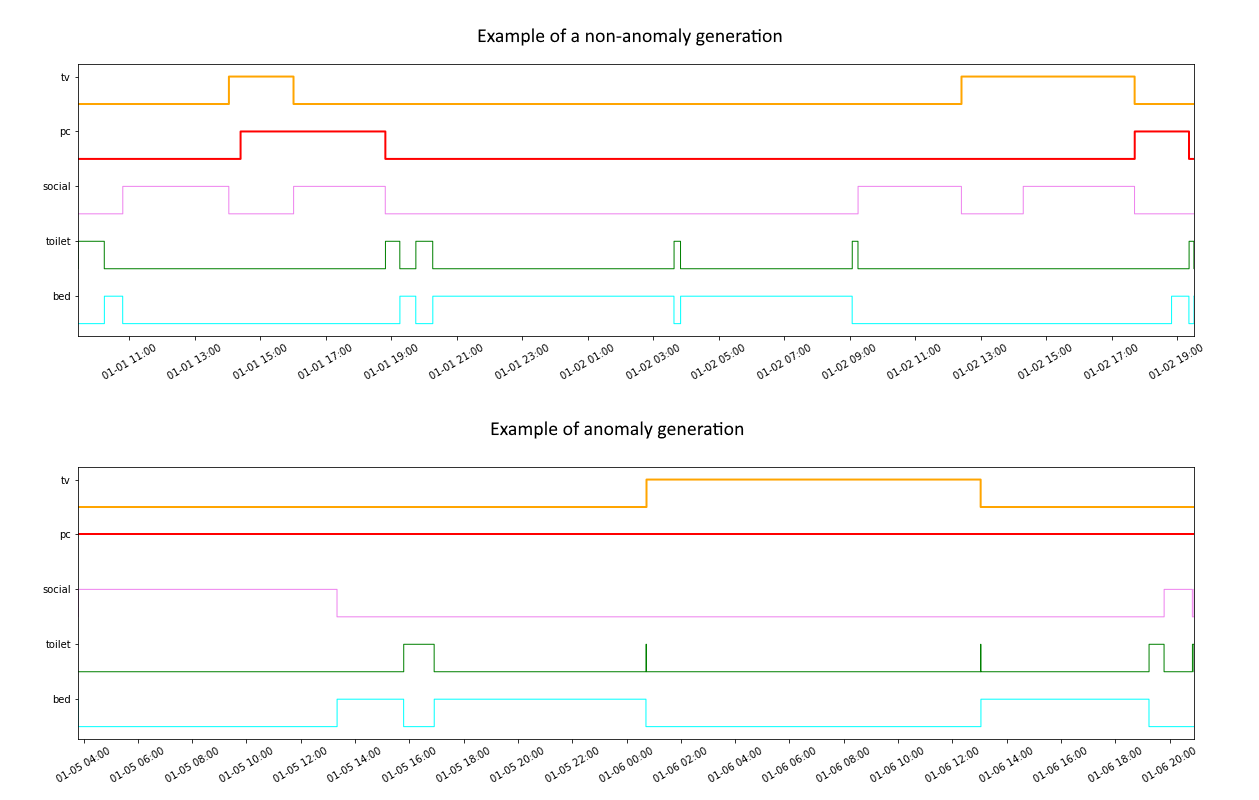}
  \caption{Plotted generated dataset.}
  \label{fig:plot}
\end{figure}
 
 Figure 6 shows two plots of two different portions of the data generated by the generator\_from\_matrix object, with each plot having five different binary signal, one for each sensor. In the first plot we can see the normal behavior of the emulated subject, where the durations of the sensors mirrors the values in the input files and correctly emulate a common person lifestyle, while the second plot clearly shows some anomalous behaviour of the emulated subject in the period of time defined by the input anomalies file, with the subject not spending the night on the bed, leaving the pc turned on for two days, watching tv a whole night and morning and spending very little time in the toilet.

\section{Impact}
\label{}

The fact that it needs only two lines of code to generate a wide amount of data, BinarySDG permits to be used by users who don’t know R language and, being an R package, it’s easy to install on every machine that has R language installed.
Using BinarySDG, researchers working in the field of Data Science, Artificial Intelligence, Machine Learning and so on, will have a solid starting point for their studies in Smart Home Sensor Data, avoiding waiting for sensor installation in controlled environments and, of course, saving money for the installation process.
Moreover, the fact that the user can control the virtual subject’s behavior and where to generate anomalies opens possibilities for training semi-supervised anomaly detection algorithms and save time for training a completely unsupervised model.
Developing this package, our goal is to break barriers (in terms of scarcity of data, timing and costs problems) that stop students and researchers from beginning studies in this type of field, permitting more discovers and results in human behavior and Smart Home applications.

\section{Conclusions}
\label{}

BinarySDG is an innovative tool for generating synthetic sensor data whether the user has a starting dataset sample or not, allowing any kind of user to have a sample of data, of the desired size and characteristics, from where to continue his researches by simply typing two lines of R code.

\section*{Acknowledgements}
\label{}

Optionally thank people and institutes you need to acknowledge.

\section*{Required Metadata}
\label{}

\section*{Current code version}
\label{}

\begin{table}[!h]
\begin{tabular}{|l|p{6.5cm}|p{6.5cm}|}
\hline
\textbf{Nr.} & \textbf{Code metadata description} & \textbf{Please fill in this column} \\
\hline
C1 & Current code version & 1.0\\
\hline
C2 & Permanent link to code$/$repository used for this code version &  \url{https://github.com/Rocche/SensorGenerator} \\
\hline
C3 & Legal Code License   & MIT License\\
\hline
C4 & Code versioning system used & git \\
\hline
C5 & Software code languages, tools, and services used & R \\
\hline
C6 & Compilation requirements, operating environments \& dependencies & markovchain, lubridate \\
\hline
C7 & If available Link to developer documentation/manual & For example: $http://mozart.github.io/documentation/$ \\
\hline
C8 & Support email for questions & giacomorocchetti97@gmail.com\\
\hline
\end{tabular}
\caption{Code metadata (mandatory)}
\label{} 
\end{table}

\end{document}